\documentclass{elsart}
\usepackage{graphicx}
\def\be{\begin{eqnarray}}
\def\ee{\end{eqnarray}}
\begin{document}

\renewcommand{\thefootnote}{\arabic{footnote}}
\setcounter{footnote}{0}

\begin{frontmatter}
\title{\LARGE\bf Free Random L\'{e}vy Variables and Financial Probabilities}

\author[a,b]{Zdzis\l{}aw Burda},
\author[a]{Jerzy Jurkiewicz},
\author[a]{Maciej A. Nowak\thanksref{nato}},
\thanks[nato]{Talk at NATO Advanced Research Workshop
 ``Applications of Physics to Economic Modeling'', Prague, 8-10 February, 2001. }
\author[c]{G\'abor Papp},
\author[d]{Ismail Zahed}

\address[a]{\it M. Smoluchowski Institute of Physics,
Jagellonian University, Cracow, Poland}
\address[b]{\it Fakult\"at f\"ur Physik, Universit\"at Bielefeld
P.O.Box 100131, D-33501 Bielefeld, Germany}
\address[c]{\it HAS Research Group for Theoretical Physics, 
E\"otv\"os University, Budapest, H-1518 Hungary}
\address[d]{\it Department of Physics and Astronomy,
SUNY-Stony-Brook, NY 11794  U.\,S.\,A.}

\begin{abstract}
We suggest that Free Random Variables, represented here by 
large random matrices with spectral L\'{e}vy disorder, may be relevant
for several problems related to the modeling of financial systems.
In particular, we consider a financial 
covariance matrix composed of asymmetric and free
random L\'{e}vy matrices. 
We derive an algebraic equation for the resolvent and solve it to
extract the spectral density. The
free eigenvalue spectrum
is in remarkable agreement with the one obtained from the covariance 
matrix of the SP500 financial market.
\end{abstract}

\end{frontmatter}
\vskip .3cm

Noise with power law distribution is encountered 
in numerous stochastic systems in physics, biophysics and 
finances  (e.g. financial time series)~\cite{STAN,BOU}.
One of the crucial problems in such systems is an understanding 
of the character of correlations between numerous agents.
It is therefore tempting to try to find a description, 
in which we could follow the stochastic evolution  not of a single object, but 
rather an evolution of an array (matrix) built of many objects.  
Such description requires basically a generalization of the classical concepts
of theory of probability to non-commuting variables. 
This fundamental concept
has been introduced  by Voiculescu,
using the powerful theory of free random variables~\cite{VOI}. Another
major twist happened when it was realized, that abstract non-commuting
operators representing FRV have an explicit realization
in terms of  certain large random matrices, 
when the size of the matrix tends
to infinity. 
 Free random matrices with Gaussian fixed point, have been applied 
to many physical problems~\cite{GROSS,ZEE,QCD}. 
Recently, using some of the concepts developed 
in ~\cite{VOI}, we extended the concept of free random
 L\'{e}vy variables to
matrices~\cite{US00}, and suggested that the results may be relevant 
for addressing the issue of noise in stochastic systems with power
law distributions. 
In this talk, we explain in a pedagogical way what is the corner-stone
of stability for non-commuting variables. 
Then we move towards a sample application, discussing the financial 
correlation matrix.

Let us consider a Gaussian distribution in classical theory
of probability, i.e. the distribution
\be
p(x)=\frac{1}{\sqrt{2\pi}\sigma} e^{-\frac{x^2}{2\sigma^2}}
\label{gauss}
\ee
with the variance $\sigma$. Now let us ask the question, 
what is the distribution of the sum of two identical independent  Gaussian 
distributions, i.e. what is the distribution of 
\be
S_2=x_1 + x_2
\label{addgauss}
\ee
where $P_2 (S_2) = p(x_1) \otimes p(x_2)$.
Since the Fourier transform unwinds the convolution, if we introduce  
the characteristic functions
\be
\phi(q)&=&\int p(x) e^{iqx} dx \nonumber \\
p(x)&=&\frac{1}{2\pi} \int \phi(q)e^{-iqx} dq\,\,,
\ee 
then,
$\Phi_2(q) = \phi(q) \cdot \phi(q)$. The inverse Fourier transform
gives 
\be
P_2(S_2) = \frac{1}{\sqrt{2\pi}\sqrt{2}\sigma} 
e^{-\frac{x^2}{2(\sqrt{2}\sigma)^2}}
\ee
i.e. again the normal distribution but with the new variance $\sigma_2=
\sqrt{2}\sigma$. 
One may say, that the logarithm of the characteristic 
function
  \be
   \ln \Phi_2(q) = \ln \phi(q) + \ln \phi(q)
\label{lawgauss}
\ee
realizes a certain {\it additivity law} for  
Gaussian variables (\ref{addgauss}).

Is it possible to find an  analog of this construction
for random matrix ensembles?
Consider again a Gaussian  ensemble built of large $N \times N$ matrices $M$,
 for which 
the pertinent resolvent
is generically given by
\be
G(z) =  
\int dM  e^{-N\,V(M)}\frac 1N\,\, {\rm Tr}\left( \frac 1{z-M}\right)\,,
\label{0}
\ee
where the 
the potential 
$V(M)$ equals $\frac{1}{2} {\rm Tr} M^2$. For simplicity we put the scale of 
the Gaussian distribution equal to one.
This Green's function fulfills a simple algebraic equation 
$G(z)=(z-G(z))^{-1}$, and the normalizable solution reads
\be
G(z)= \frac{1}{2}(z-\sqrt{z^2-4}) \,.
\label{greeng}
\ee
Using the fact, that the averaged spectral distribution is related to the 
discontinuities of the Green's function\footnote{
It is obvious using special functions property
$\frac{1}{x+i\epsilon}={PV}\frac{1}{x}-i\pi \delta(x)$.}
\be
\rho(\lambda)= -\frac{1}{\pi} \lim_{\epsilon \rightarrow \infty} {\rm Im}
  G(z)|_{z=\lambda+i \epsilon}
\label{disc}
\ee
we recover  Wigner's semicircle law
\be
\rho(\lambda)=\frac{1}{2\pi} \sqrt{4-\lambda^2} \,.
\ee

Let us now try to {\it add} two independent matrix Gaussian ensembles, as 
we did in the case of commuting variables of the classical probability theory.
In other words, we are asking what is the distribution of the eigenvalues, 
if the corresponding Green's function is given by
\be
G_{1+2}(z) &=&
\int dM_1 dM_2  e^{-N\,V_1(M_1)}e^{-N\,V_2(M_2)}\nonumber \\ 
& & \frac 1N\,\,{\rm Tr}\left( \frac 1{z-(M_1+M_2)}\right)\, .
\label{addfrv}
\ee
This time the ``convolution'' is matrix-valued, and the
matrices $M_1,\,M_2$
do not commute!

Free Random Variable (FRV)  calculus provides a surprisingly simple solution 
to this problem. 
First, let us find the functional inverse of the Green's functions
for ensembles $M_1$ and $M_2$, i.e. find the solution $G[B(z)]=z$ for every $z$ 
for both ensembles.
Then, define $R(z)=B(z)-1/z$ for both ensembles. The function R is additive
\be
R_{1+2}(z)=R_1(z) +R_2(z) \, .
\label{frvlaw}
\ee
Proceeding now in the reverse order one can reconstruct $G_{1+2}(z)$.

We exemplify this construction considering two identical Gaussian
ensembles (\ref{0}). Since for both the Green's function is given by
(\ref{greeng}), we get
\be
B_1(z)=B_2(z)=z+1/z\
\ee
so $R_1(z)=R_2(z)=z$
and the additivity law gives
\be 
R_{1+2}(z)=R_1(z)+ R_2(z)= z+z=2z
\ee
or, equivalently, $B_{1+2}(z)=2z+1/z$. 
By substitution $z \rightarrow G_{1+2}(z)$ and by the 
definition of the functional 
inverse,  we immediately get 
\be
G_{1+2}(z)=\frac{1}{4}(z-\sqrt{z^2-8}) \, .
\ee
The discontinuity of this function yields the spectral distribution 
\be
\rho_{1+2}(\lambda)=\frac{1}{4\pi}\sqrt{8-\lambda^2} \, .
\ee
The matrix convolution of two Gaussian ensembles is again a Gaussian ensemble, 
but rescaled by a factor $\sqrt{2}$.
 
The additivity law (\ref{frvlaw}) in non-commutative probability theory
of random matrices 
forms  an  analog of the additivity law for 
the logarithms of the characteristic functions
in case of classical probability theory.

This example demonstrates that the  techniques of FRV offer a powerful 
shortcut when we seek  the distribution of variables coming from 
``sum'' of different ensembles. The above construction could be generalized
to higher-order polynomial measures, 
sums of random and deterministic ensembles, 
free products of ensembles and also to strictly non-hermitian
 ensembles~\cite{ZEE,QCD}. 

We focus now on the crucial question of stability.
We know that in the classical probability 
theory the concept of Gaussian stability
(central limit theorem) was generalized by L\'{e}vy and Khintchine to 
distributions with a power-law tail. 
Is it possible that the Gaussian stability for FRV allows for a
 similar generalization?
Works of Bercovici and Voiculescu~\cite{VOI} provide a positive, but formal
answer to this problem. Recently, we have suggested an explicit construction
for the matrix measure of such ensembles, borrowing on the Coulomb gas analogy
well-known in random matrix theory. Generically, the potential is given by 
\be
V'(\lambda)=2\, \mbox{\rm Re }\, G(\lambda +i0)\,\,.
\ee
and for free L\'{e}vy matrices,  $G(z)$ satisfies an algebraic equation
in large $N$ limit~\cite{US00} ($\alpha\neq 1$)
\be
\label{4}
b\,G^{\alpha}(z)-(z-a)\,G(z)+1=0\,\,,
\ee
in the upper half-plane, and follows by Cauchy reflection in the lower
half-plane. The parameter $b$ is related to the L\'{e}vy index $\alpha$, 
asymmetry $\beta$, and range or scale  $\gamma$. Further  details 
are  discussed 
in~\cite{US00}. We would like to mention, that the explicit form 
of the matrix measure is known only in few cases.
Even in these cases, L\'{e}vy ensembles exhibit very nontrivial
behavior, and the potentials are non-analytic, contrary to the
broadly studied polynomial cases.

Let us apply some of these observations to econophysics. 
Recently, it was pointed out that financial covariances are
permeated by Gaussian noise in the low-lying eigenvalue region 
with consequences on risk assessment~\cite{RAN}. Here we show,
following our recent analysis~\cite{USII}, that 
the financial covariance spectrum is throughout permeated by free L\'{e}vy noise.
We quantify the fluctuations in a 
stochastic system by the use of
a covariance matrix ${\bf C}$,
\be
{\bf C}_{ij} =\frac 1{T^{2/\alpha}} \,\,\sum_{t=1}^T\,\,
M_{ti}\,M_{tj}\,\,,
\label{5}
\ee
where $M_{ti}$ is $T\times N$ (asymmetric).
The normalization in (\ref{5})
follows from the fact that the ``variance'' for L\'{e}vy matrices grows like
$T^{2/\alpha}$ where $\alpha=2$ is the expected diffusive limit (Brownian).
Free L\'{e}vy ensembles are super-diffusive for $0<\alpha<1$ and
sub-diffusive for $1<\alpha<2$ with $\alpha=1$ the critical divide.
We define  the empirical matrix of relative returns
\be
M_{ti} = \left( x_{it+1}-x_{it}\right)/x_{i0} - \langle M\rangle\,\,,
\label{M1}
\ee
where $x_{it}$ are the raw returns made of the price $x_{it}$ 
of stock i at time t. The returns are normalized to the initial
price $x_{i0}$ to insure scale invariance. They carry zero mean
after subtracting the average relative return 
$\langle M_{ti}\rangle_{t,i}$. 
For the raw
prices we will use the daily quotations of $N=406$ stocks 
from the SP500 market over the period of $T+1=1309$ days 
from 01.01.1991 till 06.03.1996 (ignoring dividends). For these
data the matrix asymmetry is $m=1308/406\approx 3.22$.

The spectrum of ${\bf C}$ contains important information about the
character of the correlations. 
Using Voiculescu's
powerful machinery of $R$ and $S$ transforms for free random 
variables~\cite{VOI} and some of the techniques developed in our earlier papers
 we find that
the resolvent for the covariance built from pure L\'{e}vy matrices $M$
 satisfies the transcendental equation ($\gamma=1$) \cite{USII}
\be
G(z)=\frac{1+w(z)}{z} \,,
\ee
where $w(z)$ satisfies the multi-valued equation
\be
-e^{i \frac{2\pi}{\alpha}} \cdot w^{2/\alpha} \cdot z =(1+w)(w+m)\,\,,
\ee
for fixed asymmetry $m=T/N$ and  L\'evy asymmetry parameter $\beta=0$.
The distribution of eigenvalues of the covariance matrix follows from
the discontinuity of $G(z)$.
The distribution is unimodal
and it's support extends to infinity. Since the covariance is a square, the
tail distribution is characterized by an index $\alpha/2$ since
the entries $M_{ti}$, are L\'{e}vy distributed with index $\alpha$. 

\begin{figure}
\centerline{\includegraphics[width=0.8\textwidth]{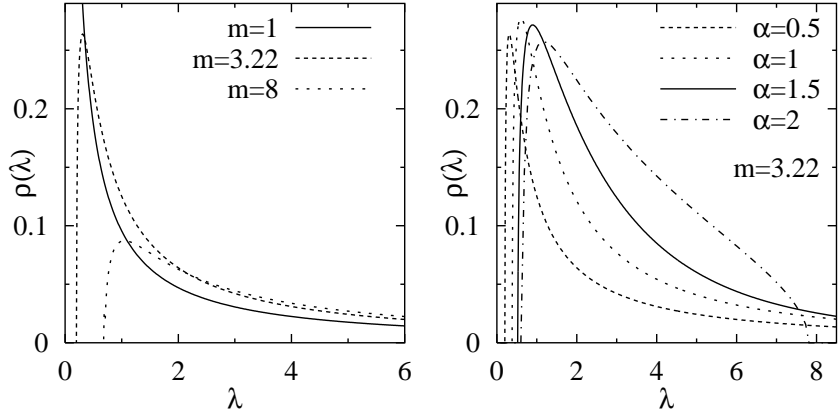}}
\caption{
Left: Spectral density of FRV with  $\alpha=1/2$ and different 
asymmetry parameters $m$. Right: spectral density for 
several indices $\alpha$, at $m=3.22$. 
}
\label{fig2}
\end{figure}

 In Fig.~\ref{fig2} 
we show some typical spectra. The
case $\alpha=2$ corresponds to 
Gaussian distributions used recently in ~\cite{RAN}.

\begin{figure}
\centerline{\includegraphics[width=0.8\textwidth]{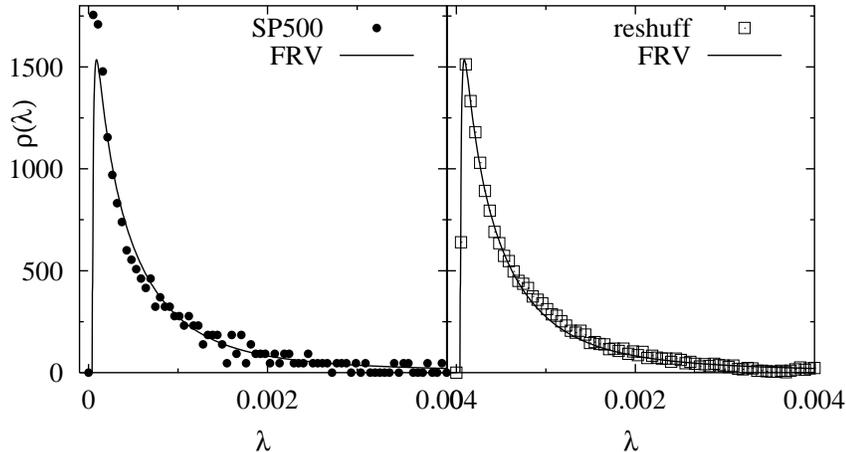}}
\caption{ 
Raw (left)  and reshuffled  (right) SP500
spectrum compared to the FRV result with $m=2$.
}
\label{fig3}
\end{figure}

In Fig.~\ref{fig3} we compare the analytical results following from the free 
L\'{e}vy covariance to the ones following from the SP500 market. 
Fig.~\ref{fig3}  shows
the distribution of eigenvalues for free L\'{e}vy matrices with 
index $\alpha/2=3/4$ and asymmetry $m=2$, versus the raw SP500 
data (left) and the reshuffled SP500 data (right). 
The reshuffled data are obtained by randomly permuting the time ordering
of the price series, independently  for all stocks, destroying all
inter-stock correlations. Our optimal fit preserves the index of
the cumulative distribution, albeit for a smaller asymmetry ($m=2 <3.22$).
It is remarkable that our free L\'{e}vy fit to the spectrum of the covariance
suggests that the time series of returns is power law distributed with index
$\alpha=3/2$ which is consistent with detailed studies of the SP500
financial time series~\cite{BOU}.

In this talk, we have summarized some of our recent results
on free random L\'{e}vy matrices and finances~\cite{US00,USII}.
A more thorough analysis and discussions including (non-rotationally
invariant) ensembles of L\'{e}vy matrices~\cite{BC} can be found in
\cite{US00,USII}. We have shown that the spectral density of the
SP500 financial covariance is in overall agreement with a free
L\'{e}vy distribution of index $\alpha/2=3/4$. Free L\'{e}vy noise 
may be dominant in financial covariances, a point that may affect
all appraisals of risk in finances. In general, we expect the
results discussed in this talk to be of relevance to a number of
problems involving non-commutative probabilities, whether in
physics, biophysics, social sciences or finances.

\vskip 1cm
{\bf Acknowledgments:}
\vskip .5cm
MAN  thanks the organizers of the Workshop for 
their hospitality.
This work was supported in part by the US DOE grant DE-FG02-88ER40388,
by the Polish Government Project (KBN)  2P03B 01917, by the Hungarian
Ministry of Education FKFP grant 220/2000 and by EC IHP grant 
HPRN-CT-1999-00161.

\end{document}